# Effect Of Slope Angle On The Runout Evolution of Granular Column Collapse for Varying Initial Volumes

**Qiuyu Wang**, Reihaneh Hosseini, and Krishna Kumar, Department of Civil, Architectural and Environmental Engineering, The University of Texas at Austin, Austin, TX 78712; email: wangqiuyu@utexas.edu

ABSTRACT: In nature, submarine slope failures usually carry thousands of cubic-meters of sediments across extremely long distances and cause tsunamis and damages to offshore structures. This paper uses the granular column collapse experiment to investigate the effect of slope angle on the runout behavior of submarine granular landslides for different initial volumes. A two-dimensional coupled lattice Boltzman and discrete element method (LBM-DEM) approach is adopted for numerically modeling the granular column collapse. Columns with four different slope angles and six different volumes are modelled under both dry and submerged conditions. The effects of hydrodynamic interactions, including the generation of excess pore pressures, hydroplaning, and drag forces and formation of turbulent vortices, are used to explain the difference in the runout behavior of the submerged columns compared to the dry columns. The results show that at any given slope angle, there is a threshold volume above which the submerged columns have a larger final runout compared to their dry counterpart, and this threshold volume decreases with slope angle.

## 1 INTRODUCTION

Submarine landslides have destructive runout potential capable of transporting large volumes of sediments (100,000 m$^3$) over long distances (100 km), even at extremely small inclinations (1°) (Korup et al. 2007, Harbitz et al 2014). The runout behavior of granular material is often investigated using a small-scale granular column collapse experiment, which reveals the flow dynamics in common with large landslides (Straub 1997, Lajeunesse et al. 2005, Staron & Lajeunesse 2009). The granular column collapse experiment has been widely used to study various factors influencing the runout behavior, such as initial aspect ratio and packing density of the granular column, grain shape, slope angle, and fluid characteristics (Lube 2004, Thompson & Hupper 2007, Rondon et al. 2011, Topin et al. 2012, Pailha et al. 2013, Kumar et al. 2017, Bougouin & Lacaze 2018).

Previously (Wang et al. 2020), we numerically studied the effect of volume on the runout behavior of submerged granular columns compared to subaerial scenarios. We found that for smaller volumes, the hydrodynamic interactions inhibit the granular flow, resulting in smaller runouts compared to the subaerial counterparts, while for larger volumes, these interactions assist the granular flow, resulting in larger runouts. In our previous study, we analyzed the runout behavior of granular column collapse on a horizontal plane; however, to incorporate the hydrodynamic interactions for a realistic landslide topography, we need to take into account the effect of slope angle. In this study, we numerically examine the combined effects of volume and slope angle on the runout behavior of collapse of submerged granular columns using a two-dimensional (2D) coupled lattice Boltzman and discrete element method (LBM-DEM) approach.

The runout of a granular column collapse develops in three consecutive stages: initiation, spreading and settlement, each of which is subject to different hydrodynamic interactions. The presence of water can either enhance the runout through hydroplaning (entrainment of water between the flow front and the base) and lubrication or inhibit the runout through dilatancy-induced negative pore pressure generation and viscous drag forces (Legros 2002, Paiha et al. 2008, Lucas et al.2014, Kumar et al. 2017). In the remainder of this paper, after a brief description of the numerical method and simulation setup, we describe the three stages of granular collapse in more detail, followed by a closer look into the individual hydrodynamic interactions that predominate each of the stages and their effects on the runout behaviour of granular columns for different volumes and slope angles.

## 2 NUMERICAL METHOD

For this study, we use a 2D coupled lattice Boltzman and discrete element method (LBM-DEM) approach, similar to our previous study (Wang et al. 2020). The LBM models the fluid flow modeled at the mesoscopic scale, while DEM is used to capture the interactions of individual soil grains. By performing a momentum exchange between the interstitial fluid and soil grains, we study the hydrodynamic interactions and fluid flow behavior at the pore scale. To approximate the real 3D flow condition in a 2D simulation, we adopt a hydrodynamic radius, in which we reduce the radius of the grains by 20% (Boutt et al. 2007) only for the LBM computations. For further information on the coupling between the LBM and DEM, refer to Kumar et al. (2017) and Wang et al. (2020).

## 3 SIMULATION SETUP

We investigate the runout behavior of granular columns with six different initial volumes, 10,000, 20,000, 30,000, 40,000, 50,000, and 60,000 cm$^3$, and for four different slope angles, 0°, 2.5°, 5°, and 7.5°, at each volume. The slope angle, θ, is accounted for by changing the direction of gravitational acceleration, g, as shown in Figure 1. For all the columns, the aspect ratio, defined as the ratio of the initial height ($H_i$) to the initial length ($L_i$), is 0.2.

The columns are constrained by a wall on one side and a gate on the other as shown in Figure 1. The initial configuration is created by ballistic deposition of polydisperse grains between the wall and the gate in the absence of fluid, resulting in an average initial packing density (the ratio of the volume of solids to the total volume) of 84%. The grains are modelled as discs with radii between 0.5 and 0.9 mm, density of 2650 kg/m$^3$, contact friction angle of 28°, and linear spring stiffness of $1.6 \times 10^6$ N/m. The coefficient of restitution, which controls the energy lost during the collision of grains, is 0.26 in this study. Once the grain deposition reaches equilibrium in the dry condition, the surrounding fluid is enabled to create a submerged column at equilibrium, and the gate is instantaneously released to cause the collapse. The final runout distance ($L_f$) is measured at the furthest grain with at least three contacts to the main mass to avoid a runaway grain affecting the runout distance.

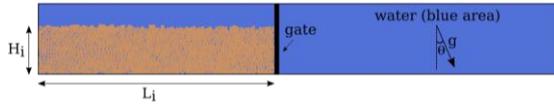

Figure 1. Configuration of underwater granular column collapse

## 4 EVOLUTION OF A GRANULAR COLLAPSE

The runout evolution of a granular column collapse involves three distinct stages: (a) initiation, (b) spreading, and (c) settlement. Figure 2 shows the snapshots of the runout evolution of the largest volume granular column (60,000 cm$^3$) on a slope of 7.5°. In all the figures in this study, we have normalized the time by a characteristic time, defined as $\tau_c = \sqrt{(H_i/g)}$ (Lajeunesse et al. 2005). The *initiation stage* is characterized by the mobilization of grains above a shear plane, which originates from the toe of the column and is inclined at 45-50° with the horizontal, illustrated as AA' in Figure 2a. During the initiation stage, most of the initial potential energy is converted to kinetic energy. Subsequently, in the *spreading stage* the mobilized kinetic energy is dissipated by the horizontal spreading of the grains. This stage is characterized by the interaction between the spreading mass and the surrounding fluid, resulting in hydroplaning at the bottom, viscous drag at the flow front, and the formation of eddies at the top surface (see Figure 2c and 2d). As the initial volume increases, the number of eddies increases proportionally due to a proportional increase in the interacting surface area. The final stage is the *settlement phase*, which features the slow down and stoppage of the spreading granular mass. As the granular mass loses horizontal velocity, the eddies developed during the spreading stage start to depart from the free surface (Figure 2e).

Figure 3 shows the evolution of normalized runout distance (($L_f$ - $L_i$)/$L_i$) with time for different initial volumes (10,000 to 60,000 cm$^3$) and varying slope angles in dry and submerged cases. In general, the submerged cases have a lower rate of runout evolution, but flow for a longer amount of time. Although the three distinct stages of runout can be observed for all slope angles, the total duration increases as the slope angle increases.

Figure 4 summarizes the final normalized runout for both submerged and dry cases at different slope angles and volumes, while Figure 5 shows the difference between the submerged and dry final normalized runouts. The normalized final runout increases with volume and slope angle for both submerged and dry case. For the submerged cases, the rate of increase in the final runout with each increment of slope angle is higher for larger volumes, while for the dry cases, this rate seems to be independent of the volume. This observation is made based on the change of the slopes of the lines in Figure 4. We had previously (Wang et al. 2020) observed that, at a slope angle of zero, for the largest volume, the normalized final runout distance of the submerged column was higher than the dry column, while for all the other smaller volumes, the runout of submerged column was less than the dry column. In this study, we observe a similar trend in the final runout for non-zero slope angles, except that the threshold volume at which the submerged granular flow runs further than its dry counterpart decreases non-linearly as the slope angle increases, as shown by the red curve in Figure 5.

In the following sections, we discuss different hydrodynamic interactions that contribute to the observations made above.

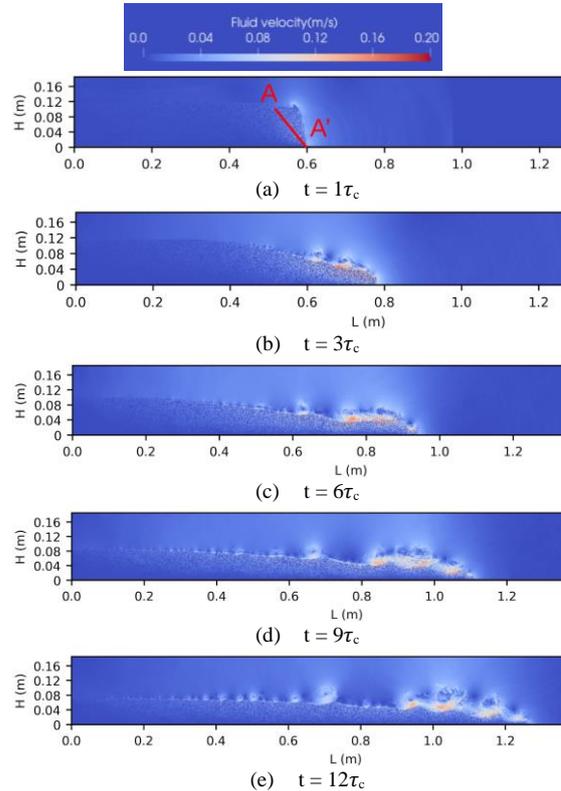

Figure 2. Flow evolution of a granular column collapse in fluid (initial aspect ratio a = 0.2, slope angle θ = 7.5°, volume = 60,000 cm$^3$)

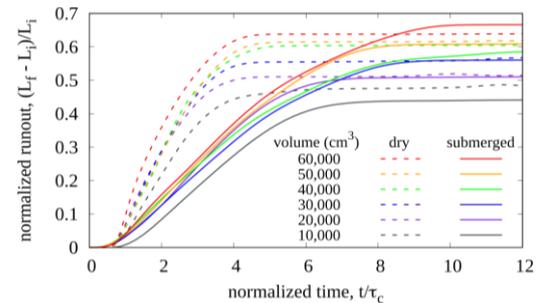

(a) θ = 0°

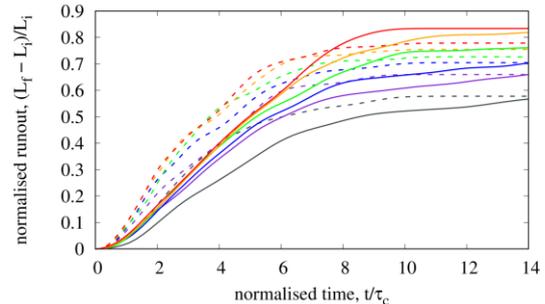

(b) θ = 2.5°

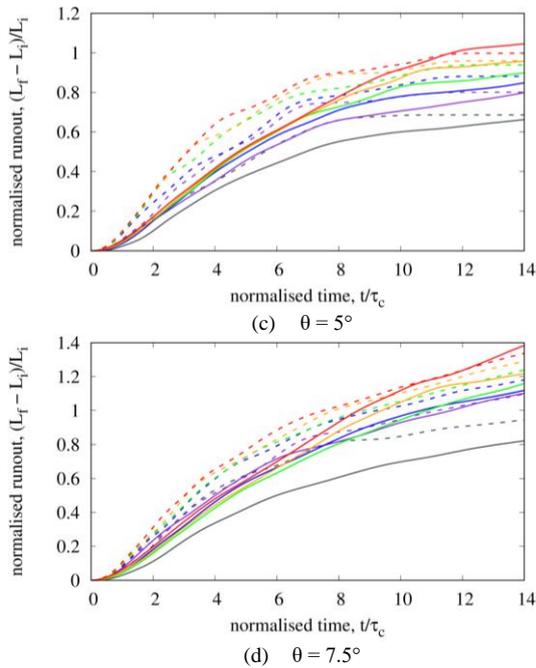

Figure 3. Evolution of normalized runout distance with time for dry and submerged granular columns with varying initial volumes and slope angles.

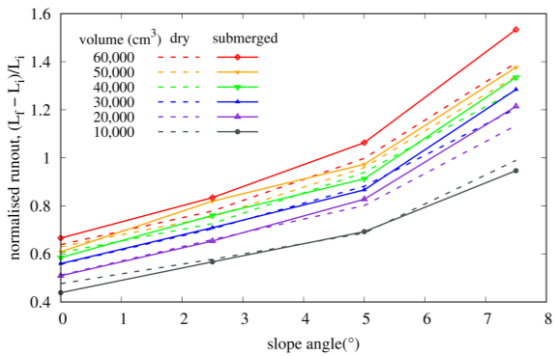

Figure 4. Final normalized runout versus slope angle, for different initial volumes.

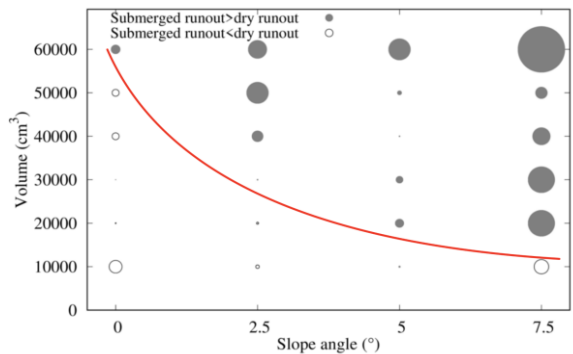

Figure 5. Final normalized runout difference between dry and submerged cases versus slope angle, for different volumes. The size of the bubble shows the magnitude of the difference in the normalized runout distances.

### 4.1 The generation of negative pore pressure

During the initiation stage of a submerged granular collapse, as the initially dense granular material is sheared along the failure surface, dilatancy-induced negative pore pressures develop. The development of these negative pore pressures results in a slower runout evolution in submerged cases compared to the dry cases. Figure 6 shows the measured negative excess pore pressures along the shearing plane at $t = 1\tau_c$, a time during the initiation stage, for different initial volumes and slope angles.

Although there is no single uniform trend in terms of the change of negative excess pore pressure with slope angle and volume, the general trend seems to be a decrease of negative excess pore pressure with increasing slope angle as well as increasing volume. The trend regarding the slope angle can explain why the rate of runout in the initiation stage for submerged and dry cases becomes more similar as the slope angle increases, observed by the rate of runout evolution in Figure 3. In addition, since the decrease of negative excess pore pressure means lower inhibitive effect on runout evolution, this trend can partially explain the reduction of the threshold volume with increasing slope angle, as seen in Figure 6.

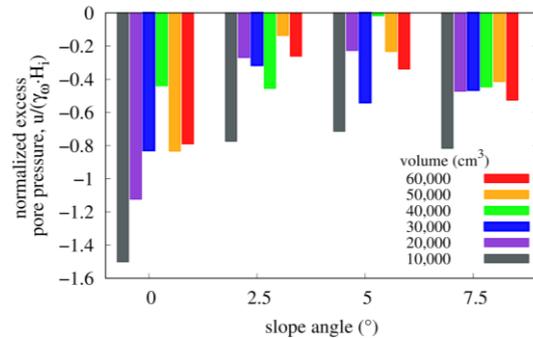

Figure 6. Normalized average excess pore pressures along the failure plane at collapse initiation for different volumes at $\tau_c$. The pore pressure ($u$) is normalized by the initial maximum pore pressure ($\gamma_\omega H_i$) and averaged over the length of the shearing plane.

### 4.2 The occurrence of hydroplaning

The runout develops most rapidly during the spreading phase. For both dry and submerged granular collapse, the kinetic energy mobilized during the initiation stage is dissipated primarily by basal friction and inelastic collisions in the spreading stage (Legros 2002). However, in the submerged cases, this basal frictional dissipation is mitigated by the water entrainment. The existence of water entrainment can be assessed by comparing the vertical effective stress at the base of the flow for submerged and dry cases. As shown in Figure 7, as the slope angle increases, the effective vertical stress for the submerged decreases substantially in comparison to the stresses in the dry cases, indicating a pronounced effect of water entrainment.

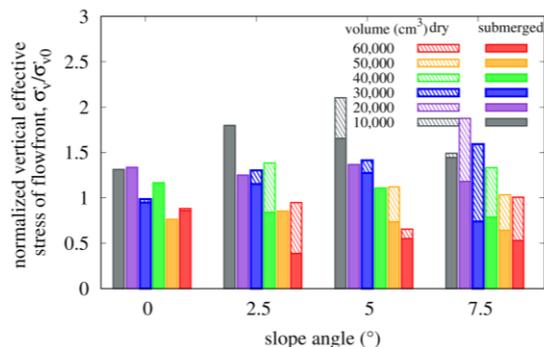

Figure 7. The vertical effective stress (contact stress) of flow front (~15 grain diameters from the furthest grain) versus initial volumes at $4\tau_c$ for different inclinations

The loss of frictional resistance due to the entrainment of water at the flow front is called *hydroplaning*. Harbitz (2003) observed that hydroplaning is most likely to occur as the densimetric Froude's number reaches the critical value of 0.4. Froude's number is defined as the ratio of flow inertia to gravity and is expressed as $F_{rd} = U/\sqrt{\frac{\rho_d}{\rho_\omega}gH}$ where $U$ is the average velocity of the sliding mass at flow front, $\rho_d$ and $\rho_\omega$ are the density of soil and water, respectively, $H$ is the average thickness of the flow front, and $g$ is the gravitational acceleration. $F_{rd}$ is useful to quantify the likelihood of occurrence of hydroplaning. With an increase in volume, the thickness of the flow front H increases, similarly, as the potential energy increases with volume, the average velocity of the flow front $U$ also increases. Hence, $F_{rd}$ will increase as $U$ is divided by the square root of $H$, indicating a higher likelihood of hydroplaning for larger volumes. Figure 8 shows the evolution of Froude's number for different volumes. We had previously observed (Wang et al. 2020) that the possibility of hydroplaning increases with increasing column volume. In this study we observe that this possibility also increases with slope angle, as seen by an increase in the magnitude of Froude's number. For example, for 10,000 cm³, the peak value of Froude's number fails to reach 0.4 on the horizontal plane but exceeds 0.4 on the sloped plane. In addition, the duration of hydroplaning increases with slope angle. For example, the duration of possible hydroplaning for 60,000 cm³ is about $6\tau_c$ at the slope angle of 0°, while this duration increases to $14\tau_c$ for the slope angle of 7.5°. As the slope angle and volume increases, the likelihood of hydroplaning in submerged cases also increases thus decreasing the dissipation through basal frictional dissipation, contributing to a larger runout compared to the dry counterparts.

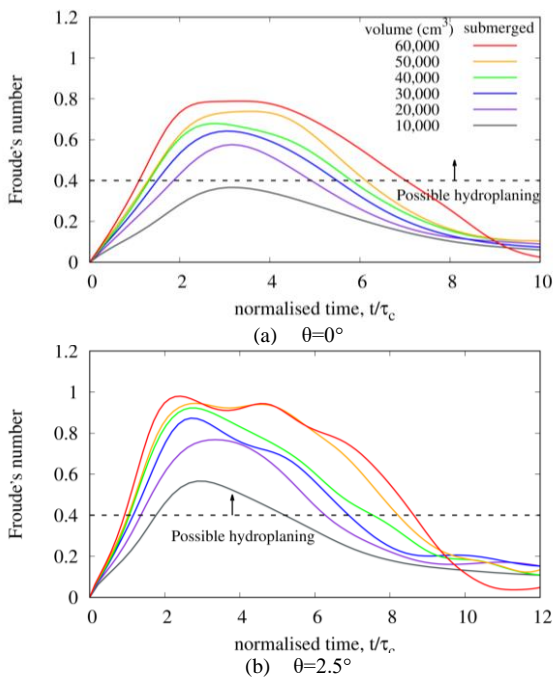

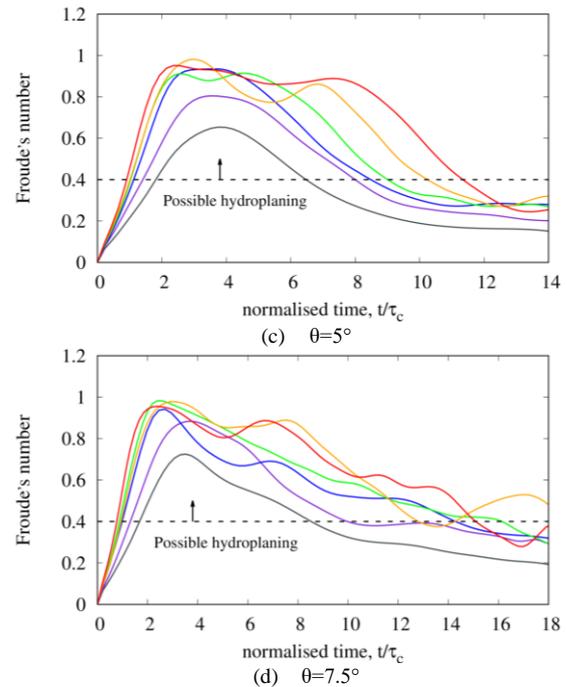

Figure 8. The evolution of Froude's number with time for different volumes

*4.3 The effect of drag forces and turbulent vortices*

During the spreading and settlement stages of the submerged collapse, energy is partially dissipated by the drag forces at the flow front and the formation of turbulent vortices along the granular surface. The effect of drag forces and turbulent vortices are compared by quantifying the hydrodynamic forces at the flow front. The hydrodynamic force ($F_{hydro}$) is normalized with respect to the total gravitational force ($F_G$) acting on the flow front. The dissipated kinetic energy depends on both the hydrodynamic forces and the free surface area (SA) on which the drag forces and turbulence act. Figure 9 shows the peak normalized hydrodynamic force (over the entire duration) with respect to the normalized free surface area. The free surface area is normalized by volume of the collapse column (V). At a given slope angle, as the volume increases, the normalized hydrodynamic force increases while the normalized surface area decreases; therefore, the total effect of drag forces remains rather constant. At a given volume, as the slope angle increases, the normalized hydrodynamic force and the normalized surface area both increase, showing a higher effect of drag forces. The slope angle has more impact on the hydrodynamic forces for larger volumes.

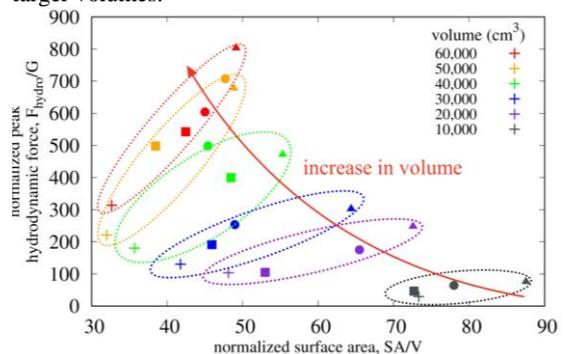

Figure 9. The peak normalized hydrodynamic force with respect to the normalized free surface area for different volumes and slope angles (cross: θ = 0°, square: θ = 2.5°, circle: θ = 5°, triangle: θ = 7.5°)

## 5 SUMMARY

Two-dimensional LBM-DEM simulations of dry and submerged granular columns were conducted to assess the effect of slope angle on the runout behaviour of granular columns for varying initial volumes. The differences in the runout behaviour between dry and submerged collapses are caused by hydrodynamic interactions, including the generation of excess pore pressures, hydroplaning, and drag forces and formation of turbulent vortices. Firstly, the generation of dilatancy-induced negative excess pore pressures during the shearing of the column in the initiation stage slows down the runout evolution in the submerged cases compared to the dry. Secondly, hydroplaning reduces the frictional dissipation at the base of the flow, therefore allowing a longer duration of runout and potentially larger final runout in the submerged cases compared to the dry. Finally, drag forces and turbulent vortices have the opposite effect compared to hydroplaning as they dissipate the kinetic energy of the granular flow, resulting in a smaller final runout for the submerged cases. Whether a submerged column will have a larger or smaller runout compared to its dry counterpart depends on which hydrodynamic interaction is predominating for that specific column volume and slope angle.

In terms of rate of runout, we observed that regardless of the slope angle and volume, the submerged cases have a lower rate of runout evolution compared to the dry cases, however, the difference between the submerged and dry runout rates decreases with increasing slope angle, due to the reduction in negative excess pore pressure. In terms of the final runout, we recognized that for each slope angle, there is a threshold volume above which the submerged collapse flows further than its dry counterpart. In other words, below this threshold, the inhibiting effect of drag forces and the negative pore pressures predominate the effect of hydroplaning, while above the threshold the assisting effect of hydroplaning overweighs the effect of other hydrodynamic interactions. Finally, we found that the threshold volume decreases with increasing slope angle.

## 6 REFERENCES

Bougouin, A., & Lacaze, L. (2018). Granular collapse in a fluid: Different flow regimes for an initially dense-packing. *Physical Review Fluids*, 3(6).

Boutt, D. F., Cook, B. K., McPherson, B. J. O. L., & Williams, J. R. (2007). Direct simulation of fluid-solid mechanics in porous media using the discrete element and lattice-Boltzmann methods. *Journal of Geophysical Research: Solid Earth*, 112(B10).

Harbitz, C. B., Løvholt, F., & Bungum, H. (2014). Submarine landslide tsunamis: how extreme and how likely?. *Natural Hazards*, 72(3), 1341-1374.

Harbitz, C. B., Parker, G., Elverhøi, A., Marr, J. G., Mohrig, D., & Harff, P. A. (2003). Hydroplaning of subaqueous debris flows and glide blocks: Analytical solutions and discussion. *Journal of Geophysical Research: Solid Earth*, 108(B7).

Korup, O., Clague, J. J., Hermanns, R. L., Hewitt, K., Strom, A. L., & Weidinger, J. T. (2007). Giant landslides, topography, and erosion. *Earth and Planetary Science Letters*, 261(3), 578–589.

Kumar, K., Soga, K., Delenne, J. Y., & Radjai, F. (2017). Modelling transient dynamics of granular slopes: MPM and DEM. *Procedia Engineering*, 175, 94-101.

Kumar, K., Soga, K., & Delenne, J. Y. (2017). Collapse of tall granular columns in fluid. In *EPJ Web of Conferences* (Vol. 140, p. 09041). EDP Sciences.

Lajeunesse, E., Monnier, J. B., & Homsy, G. M. (2005). Granular slumping on a horizontal surface. *Physics of Fluids*, 17(10).

Legros, F. (2002). The mobility of long-runout landslides. *Engineering geology*, 63(3-4), 301-331.

Lube, Gert, Huppert, H. E., Sparks, R. S. J., & Hallworth, M. A. (2004). Axisymmetric collapses of granular columns. *Journal of Fluid Mechanics*, 508, 175–199.

Lucas, A., Mangeney, A., & Ampuero, J. P. (2014). Frictional velocity-weakening in landslides on Earth and on other planetary bodies. *Nature communications*, 5(1), 1-9.

Pailha, M., Nicolas, M., & Pouliquen, O. (2008). Initiation of underwater granular avalanches: influence of the initial volume fraction. *Physics of fluids*, 20(11), 111701.

Pailha, M., Nicolas, M., & Pouliquen, O. (2013). From Dry Granular Flows to Submarine Avalanches. In Mechanics Down Under (pp. 189–198). Springer Netherlands.

Rondon, L., Pouliquen, O., & Aussillous, P. (2011). Granular collapse in a fluid: Role of the initial volume fraction. Physics of Fluids, 23(7), 073301-073301–073307.

Staron, L., & Lajeunesse, E. (2009). Understanding how volume affects the mobility of dry debris flows. Geophysical Research Letters, 36(12).

Straub, S. (1997). Predictability of long runout landslide motion: implications from granular flow mechanics. Geologische Rundschau, 86(2), 415-425.

Thompson, E. L., & Hupper, H. E. (2007). Granular column collapses: Further experimental results. Journal of Fluid Mechanics, 575, 177–186.

Topin, V., Monerie, Y., Perales, F., & Radjaï, F. (2012). Collapse Dynamics and run-out of Dense Granular Materials in a Fluid. Physical Review Letters, 109(18), 188001.

Wang, Q., Hosseini, R., & Kumar, K. (2020). Effect of Initial Volume on the Run-Out Behavior of Submerged Granular Columns. In IFCEE 2021 (pp. 256-265).